\documentclass[twocolumn,aps,prl,amsmath,showpacs]{revtex4-1}
\usepackage{graphicx,dcolumn,bm,amssymb,amsmath}
\usepackage{color}

\usepackage{float}

\usepackage{hyperref}
\def\be{\begin{equation}}
\def\ee{\end{equation}}

\begin{document}
\title{\Large Universal origin of boson peak vibrational anomalies in ordered crystals and in amorphous materials}
\author{Matteo Baggioli$^{1,2}$ and  Alessio Zaccone$^{3,4,5}$}
\affiliation{${}^1$Instituto de Fisica Teorica UAM/CSIC, c/Nicolas Cabrera 13-15,
Universidad Autonoma de Madrid, Cantoblanco, 28049 Madrid, Spain.}
\affiliation{${}^2$Crete Center for Theoretical Physics, Institute for Theoretical and Computational Physics,
Department of Physics, University of Crete, 71003 Heraklion, Greece.}
\affiliation{${}^3$Department of Physics ``A. Pontremoli", University of Milan, via Celoria 16, 20133 Milan, Italy}
\affiliation{${}^4$Cavendish Laboratory, University of Cambridge, JJ Thomson
Avenue, CB30HE Cambridge, U.K.}
\affiliation{${}^5$Department of Chemical Engineering and Biotechnology,
University of Cambridge, Philippa Fawcett Drive, CB30AS Cambridge, U.K.}

\begin{abstract}
\noindent 
The vibrational spectra of solids, both ordered and amorphous, in the low-energy regime, control the thermal and transport properties of materials, from heat capacity to heat conduction, electron-phonon couplings, conventional superconductivity etc. The old Debye model of vibrational spectra at low energy gives the vibrational density of states (VDOS) as proportional to the frequency squared, but in many materials the spectrum departs from this law which results in a peak upon normalizing the VDOS by frequency squared, which is known as the ``boson peak".
A description of the VDOS of solids (both crystals and glasses) is presented starting from first principles. Without using any assumptions whatsoever of disorder in the material, it is shown that the boson peak in the VDOS of both ordered crystals and glasses arises naturally from the competition between elastic mode propagation and diffusive damping. The theory explains the recent experimental observations of boson peak in perfectly ordered crystals, which cannot be explained based on previous theoretical frameworks. The theory also explains, for the first time, how the vibrational spectrum changes with the atomic density of the solid, and explains recent experimental observations of this effect. 
\end{abstract}


\maketitle
Understanding the physics of vibrational excitations in condensed matter is a classical topic in modern physics~\cite{Landau,Frenkel,Pohl}, which in recent years has been largely focused on understanding the vibrational spectra of disordered systems, such as liquids, glasses and disordered crystals. In particular, a unifying framework has been sought to understand how vibrational excitations change upon going from liquid to glass, and viceversa, at the glass transition~\cite{Trachenko}. An intensively studied problem is the ubiquitous anomaly (known as boson peak) in the VDOS which appears in glasses and crystals upon normalizing the VDOS $g(\omega)$ by the Debye law $\omega^{2}$, typically at THz frequencies in atomic and molecular materials. In turn, this anomaly controls or affects all anomalies and behaviours in the specific heat, heat conduction and low-T properties of solids~\cite{Pohl}. 

It is impossible to quote all the references about experimental observations: the boson peak anomaly has been observed in oxide glasses~\cite{Monaco}, molecular glasses~\cite{Monaco2}, molecular crystals with minimal orientational disorder~\cite{Tamarit}, in polymers~\cite{Sokolov,Leporini,Palyulin}, in metallic glasses~\cite{Wang,Wilde}, in colloidal crystals with defects~\cite{Islam}, colloidal glasses~\cite{Bonn}, and athermal jammed packings~\cite{Silbert}. Importantly, however, the boson peak has been experimentally observed also in ordered single crystals with no disorder, such as molecular single crystals~\cite{Pardo2,Jezowski} and non-centrosymmetric perfect crystals such as $\alpha$-quartz~\cite{Monaco}.
The observation of a boson peak in ordered crystals is as yet unexplained since all theoretical models and approaches to the boson peak problem proposed so far assume the existence of some form of \textit{disorder} in the material.

Among those previous theories, a prominent one is the heterogeneous elasticity theory which has been developed by W. Schirmacher and co-workers~\cite{Schirmacher}. This approach uses an elegant field-theoretical scheme to derive $g(\omega)$ under the assumption that the shear elastic constant of the system is fluctuating in space according to some distribution (which may or not be Gaussian). This approach cannot explain the boson peak observed in ordered single crystals~\cite{Pardo2,Jezowski,Monaco} where the elastic constants are homogeneous and have the same value throughout the whole material. 

Other models are based on quasi-local vibrational states due to randomly-distributed soft anharmonic modes~\cite{Gurevich,Parshin}, local inversion-symmetry breaking connected with nonaffine deformations~\cite{Zaccone2011, Milkus}, phonon-saddle transition in the energy landscape~\cite{Parisi}, density fluctuations of arrested glass structures~\cite{Goetze}, and broadening/lowering of the lowest van Hove singularity in the corresponding reference crystal due to the distribution of force constants~\cite{Diezemann,Elliott} or network rigidity~\cite{Naumis}.
As mentioned above, all these approaches rely on assumptions of \textit{disorder}. 

Hence, \textit{none} of the above approaches can explain the observation of boson peak in ordered crystals~\cite{Pardo2,Jezowski,Monaco}. 

In the following we will show that no hypothesis of disorder in the system is actually needed to describe the boson peak, and that the boson peak is a hallmark of all real solids with a linear viscous damping, regardless of their internal microstructure (see also \cite{PhysRevLett.114.195502,PhysRevB.70.212301,PhysRevLett.93.245902} for previous ideas about the role of damping on the low-T thermal properties of solids). 
It will also be shown that, as emphasized in previous works, while it is true that in amorphous solids the continuum approximation breaks down at a certain length-scale (or wavevector), this is because of anharmonic dissipation coming into play. 

We start considering the following standard Hamiltonian for the anharmonic crystal~\cite{Boettger}:
\begin{equation}
H=H_{0}+H_{A}\label{eq1}
\end{equation}
where $H_0=\sum_{\lambda} \hbar\omega_{\lambda} [b^{\dagger}_{\lambda}b_{\lambda}+\frac{1}{2}]$ is the Harmonic part of the Hamiltonian and $b^{\dagger}_{\lambda},b_{\lambda}$ respectively the creation and annihilation operators. The index $\lambda$ compactly represents the pair of indices 
$(\textbf{q}j)$ where 
$\textbf{q}$ is the wavevector and $j$ is the branch index. Hence, $\omega_{\lambda}\equiv \omega_{j}(\textbf{q})$. The anharmonic part can be described, in the standard way, with terms of cubic and quartic order:
\begin{equation}
H_{A}=\sum_{n=3,4}\frac{1}{n!}\sum_{\lambda_{1}...\lambda_{n}}v \prod_{i=1}^{n}[(b_{\lambda_{i}}+b^{\dagger}_{-\lambda_{i}})].
\end{equation}
Here, $v\equiv v(\lambda_{1}...\lambda_{n})$ are coefficients related to the $n$-th order derivatives of the interatomic pair potential with respect to the lattice displacements, while the factors $[(b_{\lambda_{i}}+b^{\dagger}_{-\lambda_{i}})]$ arise upon replacing the atomic displacements with the corresponding expressions in second quantization. Hence, the above equation is nothing but the usual potential energy expansion of the lattice about the rest positions of the atoms. 

This anharmonic Hamiltonian, upon performing a standard exercise in many-body theory which can be found in all textbooks~\cite{Boettger,Khomskii,Han}, or alternatively using projection-operator methods~\cite{Lovesey}, gives rise to the following phonon Green's function for an isotropic cubic crystal~\cite{Boettger,Lovesey}: 
\begin{equation}
G_{L,T}(q,\omega)=\frac{a}{\omega^{2}-\Omega^2_{L,T}(q)+ia\,\omega\,\Gamma_{L,T}(q)},\label{Gf}
\end{equation}
where $a$ is a prefactor which depends on the choice of normalization for the bosonic creation and annihilation operators, $\Omega$ is the eigenfrequency, and $\Gamma$ the damping coefficient (which results from the imaginary part of the self-energy $\Sigma$). In this form, the eigenfrequency $\Omega$ already contains the correction due to phonon-phonon interactions. 
The above expression can be readily generalized to anisotropic crystals upon replacing $\Omega$ everywhere with $\Omega_\lambda\equiv\Omega_j(\textbf{q})$: each $j$-th branch then gives rise to an additive Green function term~\cite{Boettger}, and $\Gamma(\textbf{q})$ becomes also a function of the direction of the wavevector. In the following we specialize on the example of isotropic cubic crystals, although it is clear that the results remain qualitatively valid for anisotropic crystals.

Various choices are used in the literature for the prefactor $a$ (including e.g. $a=2\Omega_{L,T}$ of Schrieffer~\cite{Schrieffer} and $a=\Omega_{L,T}^{2}$, of Abrikosov et al.~\cite{Abrikosov}), depending on what one wants to calculate~\cite{Khomskii}. Here, since our final aim is to use the Green's function to calculate the VDOS, we use a normalization that is compatible with the dimensionality of the VDOS and we choose $a=1$. With $a=1$ our Eq. (3) coincides with the expression derived by Lovesey ~\cite{Lovesey} from the Hamiltonian Eqs.(1)-(2) above.

In the above expression, $\Gamma$ represents the phonon line-width or the phonon damping coefficient which can be measured via neutron scattering experiments. Using the microscopic approach based on the anharmonic Hamiltonian, the damping coefficient has been obtained in general form, historically, by Landau and Rumer~\cite{Landau-Rumer}. More refined calculations using cubic and quartic terms in the Hamiltonian showed~\cite{Pathak} that $\Gamma_{L,T}=D_{L,T}q^{2}$ in good agreement with experiments; hence the damping coefficient is quadratic in the module of the wavevector $q$, where $D_{L,T}$ is a constant. Hydrodynamic theories~\cite{Lubensky} as well as the Akhiezer~\cite{Akhiezer} approach (in the low-frequency regime) also provide a $q^{2}$ dependence. Importantly, the same form of Green's function Eq. (3), is also derived for disordered solids~\cite{Salamatov,Boettger}. Also in the case of amorphous solids one finds that $\Gamma \sim q^{2}$~\cite{Tanaka}. 

Therefore, the denominator of the Green's function presents poles which provide the following set of dispersion relations for transverse (T) and longitudinal (L) phonons,
\begin{align}
\omega_{L,T}&= c_{L,T}\,q-i\,D_{L,T}\,q^{2}.\label{disp}
\end{align}

The corresponding speeds of sound are given by:
\begin{equation}
c_{T}^2=\frac{\mu}{\rho}\,,\quad c_{L}^2=\frac{K\,+\,\frac{2\,(d-1)}{d}\,\mu}{\rho}
\end{equation}
where $\mu$ is the shear modulus and $K$ is the bulk modulus. In general, $c_{L}>c_{T}$ since $\mu>K$ for solids with Poisson ratio in the usual range $0<\nu<1/2$. Using Eq.\eqref{Gf} and Eq.\eqref{disp} we therefore identify $\Omega_{L,T}(q)=c_{L,T}\,q$ and $\Gamma_{L,T}(q)=D_{L,T}\,q^2$, \textit{i.e.} a diffusive-like damping as discussed above. 

With the Green's function of Eq.(3) it is now possible to calculate the VDOS. Upon considering the definition of the VDOS in terms of delta functions, together with the Plemelj identity, the following expression (see Supplementary Information for details) is recovered:\noindent
\begin{equation}
g(\omega)=-\frac{2\,\omega}{3\,\pi
\,\mathcal{N}}\sum_{q<q_D}\,\text{Im}\left\{2\,G_T(q,\omega)+G_L(q,\omega)\right\},\label{eq6}
\end{equation}
where $q_{D}$ denotes the maximum (Debye) wavenumber in the system, $q_{D}=(6\pi^2 \mathcal{N}/V)^{1/3}$ and $\mathcal{N}$ the number of atoms in the system.

This formula is useful because it allows one to calculate the VDOS $g(\omega)$ from the knowledge of the Green's functions $G_{L,T}(q,\omega)$. For our theory, the Green's functions are determined from Eqs.\eqref{Gf}, and can be built using the transport coefficients such as the shear and bulk moduli and the damping coefficients $\Gamma$, as the only input to the theory. 
Instead of using the discrete sum in Eq.\eqref{eq6} one can use an integral by means of the standard transformation $\frac{3}{q_{D}^{3}}\int_{0}^{q_{D}}q^{2}dq= \frac{1}{N}\sum_{q<q_{D}}$, but the integral is not analytical. It turns out, instead, that the series in Eq.\eqref{eq6} can be summed exactly, which leads to the following closed-form expression:
\begin{widetext}
\begin{align}
g(\omega)&=\frac{\omega}{3\,\pi\, \mathcal{N}}\,\text{Im}\Big\{\frac{1}{\omega \,\sqrt{(-c_{L}^{2}+\,i\,D_{L}\, \omega)(i \,c_{T}^2+D_{T}\omega)}}
\left[-i \sqrt{i \,c_{T}^2+D_{T}\,\omega}\left(\psi(x)-\psi(-x)+\psi(1+q_{D}+x)-\psi(1+q_{D}-x)\right)\right]\nonumber\\
&+\left[(1+i)\sqrt{-\,2 \,c_{L}^{2}+2\, i \,D_{L}\,\omega}\left(\psi(y)-\psi(-y)+\psi(1+q_{D}+y)-\psi(1+q_{D}-y)\right)\right]\Big\}
\end{align}
\end{widetext}
where $\psi$ denotes the Digamma function $\psi(z)\equiv\frac{d}{dz}\ln\Gamma(z)$ with $\Gamma$ the Gamma function, and $x=-\frac{i\omega}{\sqrt{-c_{L}^{2}+iD_{L}\omega}}$, and $y=\frac{(1+i)\omega}{\sqrt{2ic_{T}^{2}+2D_{T}\omega}}$.

\begin{figure}[htbp]
\centering
{
\includegraphics[height=5.5cm ,width=8cm]{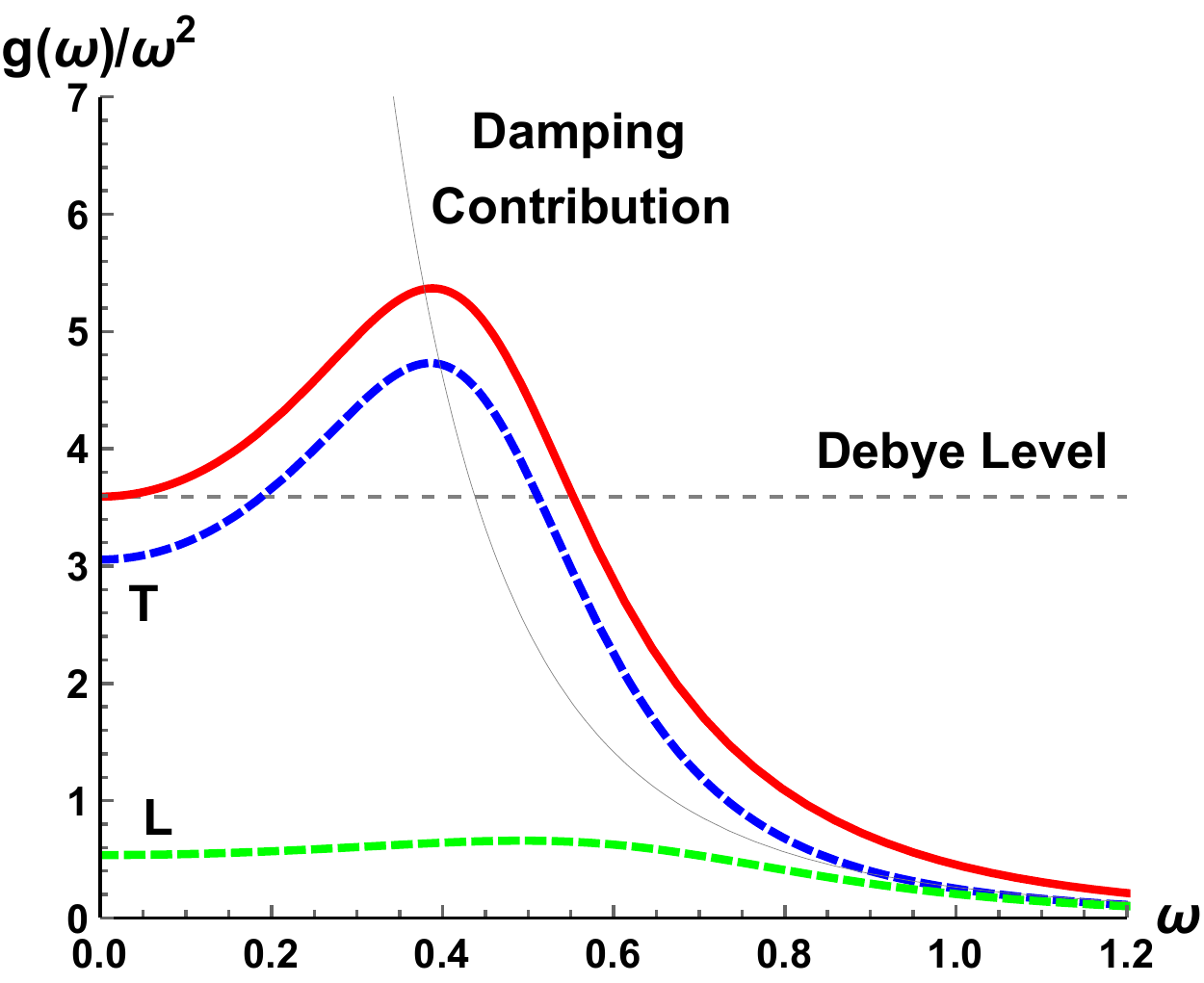}\label{Fig_1b}}\vspace{0.05cm}
{
\includegraphics[height=5.5cm ,width=8cm]{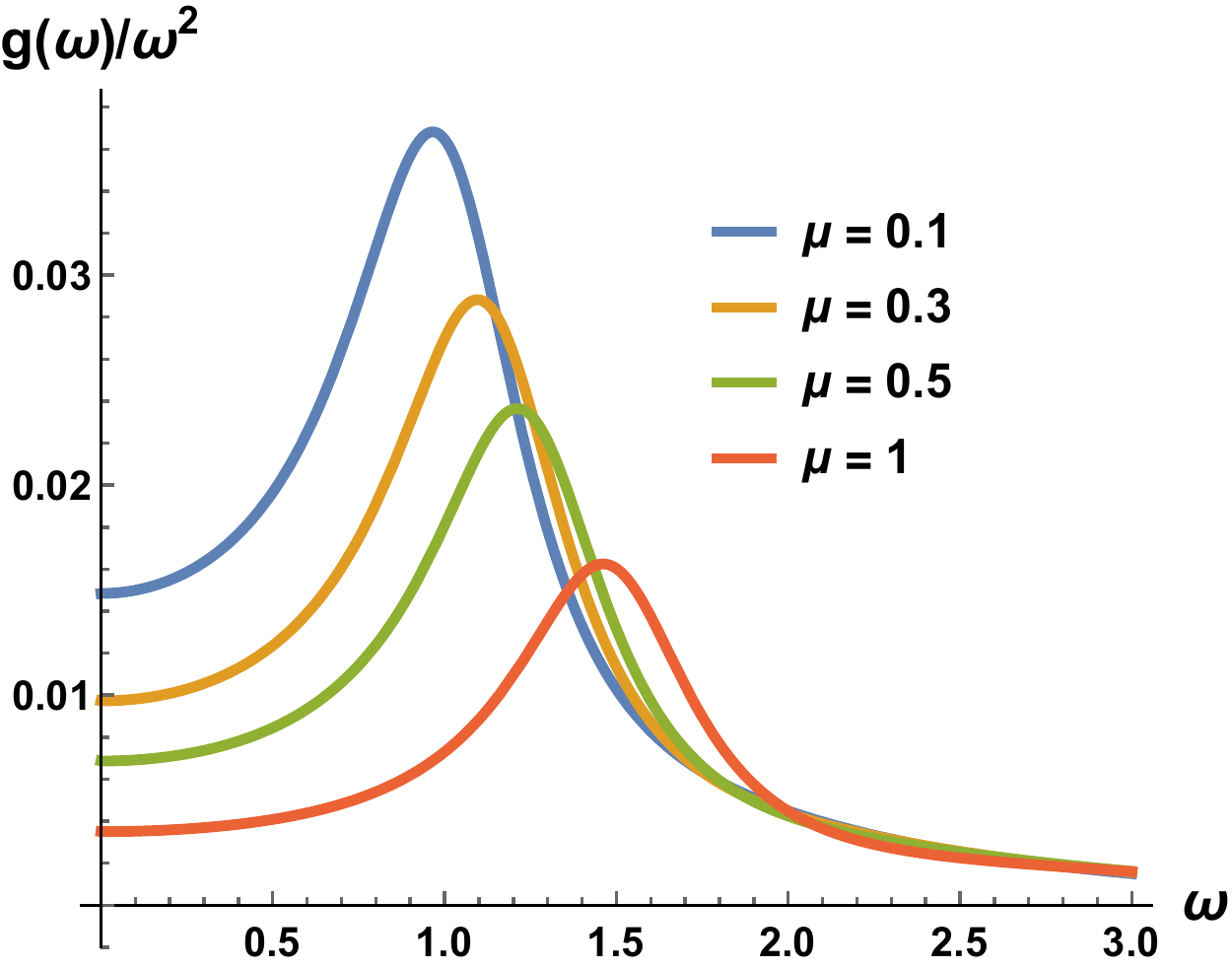}\label{Fig_1a}}
\label{Figure1}
\caption{Normalized vibrational density of states (VDOS) (a) the different contributions from propagating modes (Debye law) and dissipation are outlined along with the contributions from the transverse modes and the longitudinal ones.  In (b) as a function of the shear modulus.
}
\label{fig:Fig_1a&Fig_1b}
\end{figure}

In the following we present the predictions of the above Eq.(7) for the VDOS $g(\omega)$ at different values of the elastic constants and damping coefficients. In the normalized plots of $g(\omega)/\omega^{2}$ in Fig.\ref{fig:Fig_1a&Fig_1b}, the boson peak (BP) is evident, and results from the competition between phonon propagation (controlled by the elastic constants) and damping (controlled by the damping coefficients). The damping contribution (see gray line in Fig.\ref{fig:Fig_1a&Fig_1b}) grows monotonically upon decreasing the frequency (and it would diverge at $\omega \rightarrow 0$), and it dominates at large enough frequencies $\omega \gg \omega_{BP}$. On the contrary at small frequencies $\omega\ll \omega_{BP}$, the Debye law dominates the VDOS; from the interplay and crossover of the two contributions in Eq.(3) and (6) the boson peak is generated. This picture is in accordance with the crossover between \textit{propagons} at low frequency and \textit{diffusons} at high frequency proposed in \cite{allen1999diffusons}. In ordered crystals the diffusive-like behavior is to be attributed to phonon-phonon scattering, whereas in glasses it is mostly due to disorder, which also provides a diffusive-like damping.
Also, it is evident that the boson peak is more strongly affected by the shear elastic modulus $\mu$, and to a lesser extent by the bulk modulus $K$, which provides a theoretical basis for earlier findings of simulations~\cite{Tanaka}. In other words, the transverse contribution to the boson peak is always dominating with respect to the longitudinal one as shown in Fig.\ref{fig:Fig_1a&Fig_1b}.

\begin{figure}[htbp]
\centering
\includegraphics[height=5.5cm ,width=8cm]{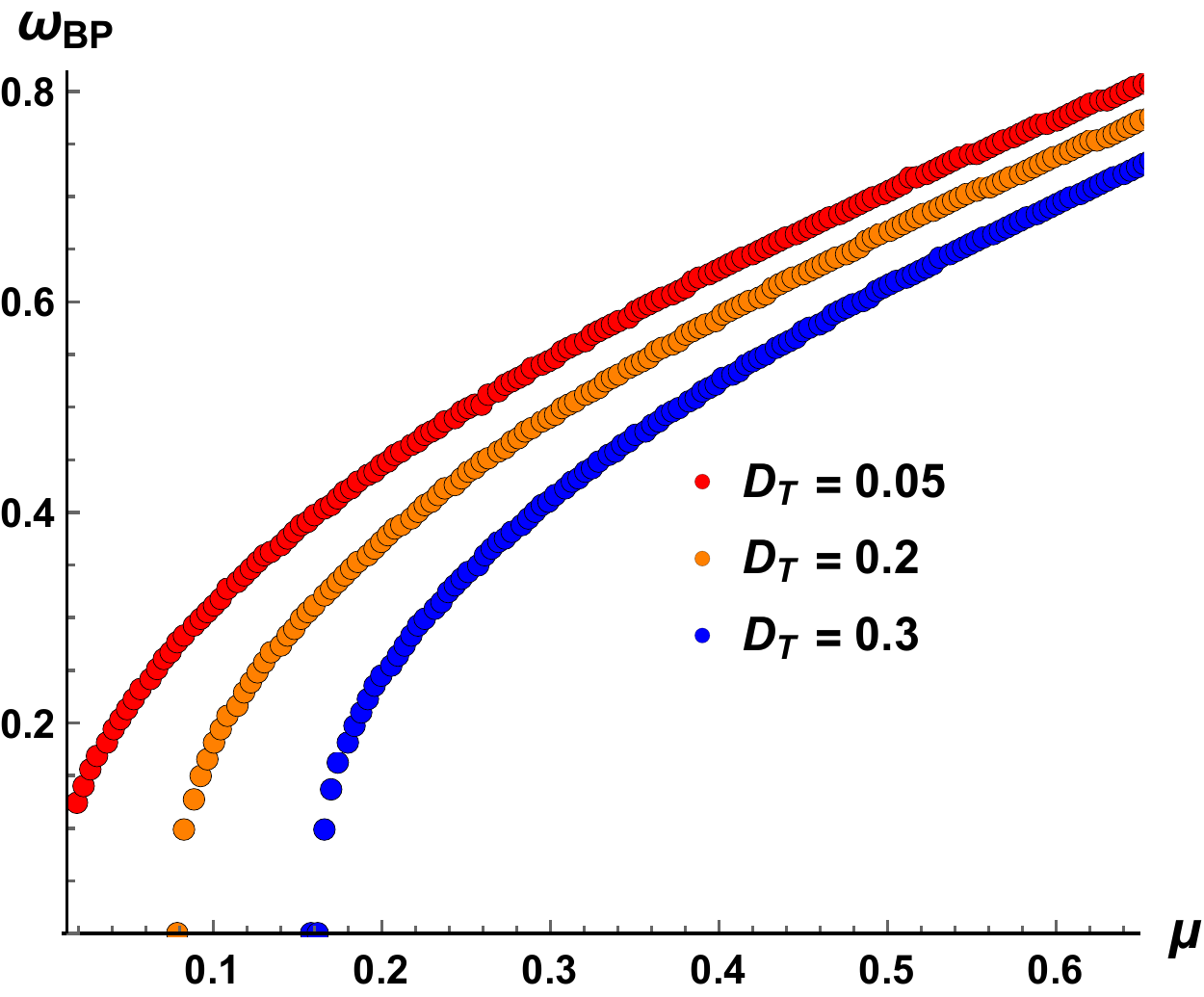}\label{Fig_2}\hfill
\label{Figure2}
\caption{Boson peak frequency as a function of the shear elastic modulus $\mu$ for different values of viscous damping. The curves are well fitted by a square root dependence.}
\label{fig:Fig_2}
\end{figure}

It is seen in Figs.\ref{fig:Fig_1a&Fig_1b} and \ref{fig:Fig_2} that upon decreasing the value of shear modulus $\mu$, the boson peak shifts to lower frequencies in a power-law (square-root) fashion, and ultimately moves to zero frequency in the limit $\mu\rightarrow 0$, which marks the limit of mechanical stability. At higher values of $\mu$ the square-root becomes more and more like a linear dependence. This prediction provides a theoretical basis to experimental results where the measured $\omega_{BP}$ decreases upon approaching the glass transition along with the vanishing of the shear modulus $\mu$~\cite{Sokolov}. A similar behaviour is observed in athermal amorphous solids~\cite{Silbert,Milkus} and athermal crystals with defects~\cite{Milkus}. 

\begin{figure}[H]
\centering
\includegraphics[height=5.5cm,width=8cm]{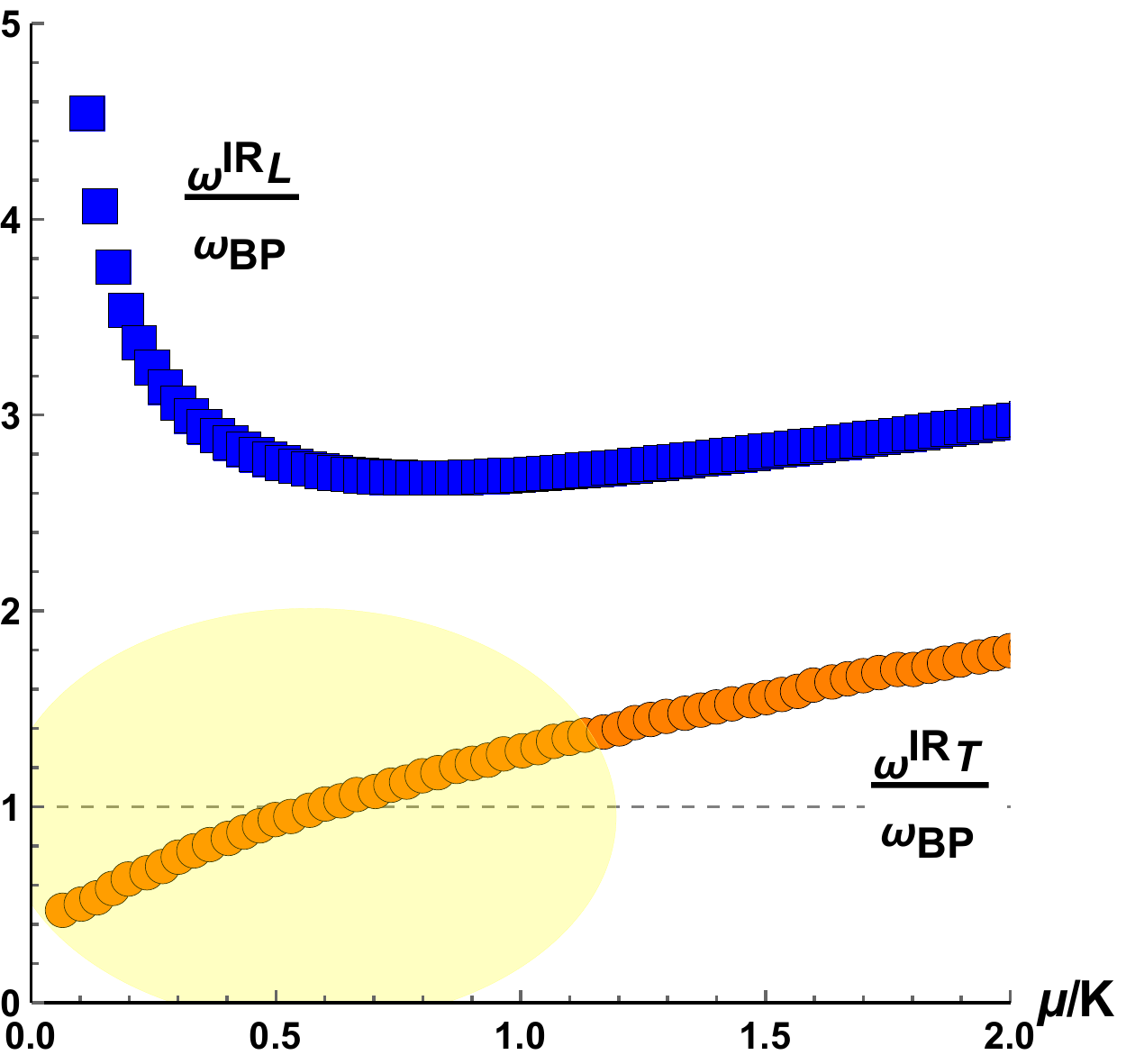}\label{Fig_4}\hfill
\label{Figure4}
\caption{Comparison between the boson peak frequency $\omega_{BP}$ and the Ioffe-Regel transverse and longitudinal frequencies $\omega^{IR}_{T,L}$ varying the ratio $\mu/K$. $D_T=0.425$ is kept fixed in the plot. The shaded area indicates the region of values $\mu/K<1$ typical of most solids, where the $\omega^{IR}_{T}/\omega_{BP}$ ratio lies in the interval $0.45-1.3$ and hence is of order unity.}
\label{fig:Fig_4}
\end{figure}

The vicinity of the boson peak frequency $\omega_{BP}$ and the longitudinal/transverse Ioffe-Regel crossover frequencies $\omega^{IR}_{L,T}=c_{L,T}^2/(\pi\,D_{L,T})$ has been investigated and discussed in \cite{Ruffle,Tanaka,beltukov2013ioffe}. The results from our analytical theory, shown in Fig.\ref{fig:Fig_4}, confirm the expectations and suggest that, especially at low damping, the boson peak frequency is very close to the Ioffe-Regel transverse frequency (but not to the longitudinal one). 
From our analytical calculations, it is clear that $\omega_{BP}$ becomes closer to $\omega_{T}^{IR}$ as the ratio $\mu/K$ gets smaller (for the majority of solids $\mu/K < 1$ which implies that the Poisson ratio lies in the usual $[0,1/2]$ interval~\cite{Landau-Lifshitz}). This provides a theoretical justification to the simulation findings of ~\cite{Tanaka} and explains the closeness of Ioffe-Regel frequency of transverse phonons to the BP frequency as due to $\mu/K<1$ in solids.

\begin{figure}[htbp]
\centering
\includegraphics[height=5.5cm ,width=8cm]{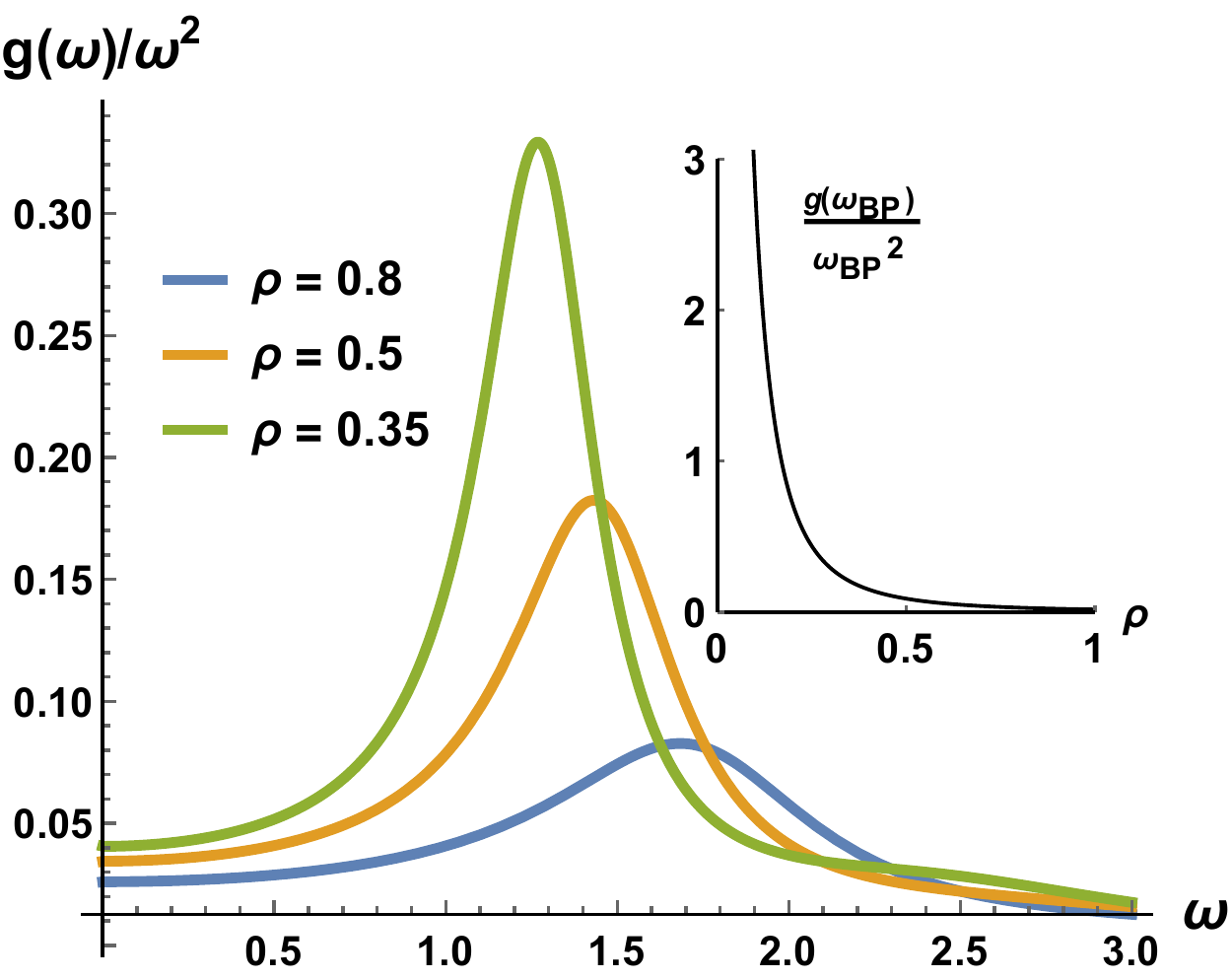}\label{Fig_5}\hfill
\label{Figure5}
\caption{The VDOS in function of the density $\rho$. The elastic moduli are taken to obey the law $\sim \alpha_1\,\rho\,+\,\alpha_2\,\rho^2$ in accordance with the experimental fits of \cite{Deschamps2014,Pabst}. For the damping $\Gamma$ we assumed a dependence $\sim \rho$ as derived in \cite{Landau-Lifshitz} for isotropic solids. The inset shows the decay of the boson peak intensity $g(\omega_{BP})/\omega_{BP}^2$ as a function of the density.}
\label{fig:Fig_5}
\end{figure}

By means of Eq.(7) one can also study the effect of atomic density $\rho=N/V$ on the boson peak and the VDOS, on the example of silica glass. In this case, importantly, the main contribution to the damping $\Gamma$ is expected to come mainly from structural disorder: hence, this example illustrates the generality of the proposed framework. We take the phonon damping $\Gamma$ to be proportional to the density $\rho$, as derived for isotropic solids in \cite{Landau-Lifshitz}, and the elastic moduli to be described by $\sim \alpha_1 \rho +\alpha_2 \rho^2$ as observed experimentally for densified silica in \cite{Deschamps2014,Pabst}. Taking into account that the Debye wavenumber $q_D$ is, by definition, proportional to the cubic root of the density $\rho$, we plot our results in Fig.\ref{fig:Fig_5}. We observe that, upon increasing the density, the intensity of the boson peak and the value of the normalized VDOS at zero frequency $\omega=0$ decrease, while the width of the peak increases and the boson peak moves to higher frequencies. To the best of our knowledge, our results represent the first theoretical explanation of the experimental data and trends on densified silica presented in \cite{Monaco}, and observed also earlier in densified $B_{2}O_{3}$~\cite{Carini}.

%\begin{figure}[tb]
%\centering
%\includegraphics[height=5.5cm ,width=8cm]{Fig_3}\label{Fig_3}\hfill
%\label{Figure3}
%\caption{Evolution of VDOS and boson peak upon varying the atomic density in the solid according to Eq. (5).}
%\label{fig:Fig_3}
%\end{figure}

In conclusion, we have provided a universal framework for the emergence of the boson peak based solely on  damping, with a focus on anharmonicity as the root cause of damping in perfectly ordered crystals, although the model is generic and also applicable to glasses where damping is, instead, mainly due to disorder. The underlying mechanism which generically produces the boson peak in the VDOS is the competition between propagation and damping or, alternatively, the coexistence of an elastic response and a viscous one determined by the damping coefficients \cite{mizuno}. 
In ordered crystals, the presence of diffusive-like damping does not rely on the existence of any disordered or amorphous structure but is caused simply by anharmonicity and phonon-phonon scattering, as shown here, which explains the observation of boson peak in ordered crystals where damping is active even in the absence of disorder, and is related to viscosity~\cite{Akhiezer,Landau-Lifshitz,Biroli}. In this way, the presented framework crucially explains the universality of the boson peak and its recent experimental observation also in perfectly ordered crystals~\cite{Pardo2,Jezowski,Monaco}.

Furthermore, this model recovers, via Eq. (7), the previously observed correlation between the Ioffe-Regel crossover frequency of transverse phonons and the boson peak frequency~\cite{Ruffle,Tanaka}.\\
\\
As a final and important result, the generic relation between VDOS and acoustic phonon damping explains the density dependence of the boson peak measured in silica glass \cite{Monaco}, for which no explanation was at hand.
Our model might be able to successfully describe also the anomalies in the thermal transport and heat capacity related to the boson peak \cite{Pohl}. In the future, it would be interesting to extend our formalism to liquids and in particular to the recently discovered gapped dispersion relations \cite{trachenko2015collective,trachenko2017lagrangian,yang2017emergence,Baggioli:2018nnp,Baggioli:2018vfc} to build a unified description of the vibrational spectra of crystals, liquids and amorphous materials. It is also interesting to notice the strong similarities between our results and the holographic models for viscoelasticity \cite{Alberte:2015isw,Alberte:2016xja,Andrade:2017cnc,Alberte:2017cch,Alberte:2017oqx,Baggioli:2018bfa}, where indeed the phonons naturally acquire a viscous damping.

All in all, this work provides new insights towards a unifying description of the vibrational spectra of solids, both ordered and amorphous.\vspace{0.5cm}

\begin{acknowledgments}
We thank A.Cano, A.Krivchikov, K. Trachenko, G. D'Angelo for several discussions and helfpul comments and J.Ll. Tamarit for reading a preliminary version of the manuscript and for comments. We acknowledge the Thomas Young Centre and the TYC Soiree where this work started.
MB acknowledges the support of the Spanish MINECO’s “Centro de Excelencia Severo Ochoa” Programme under grant SEV-2012-0249.
MB is supported in part by the Advanced ERC grant SM-grav No 669288.  
\end{acknowledgments}

\end{document}